\documentclass[a4paper]{mem}
\usepackage{natbib}
\usepackage{graphicx}
\begin{document}
   \title{Radio-optical scrutiny of the central engine in compact AGN
%\thanks{this is a place for a title footnote}
}

   \author{T.G. Arshakian \inst{1}\fnmsep\thanks{On leave from the
       Byurakan Astrophysical Observatory, Armenia}, V.H. Chavushyan
       \inst{2,3},\\ E. Ros \inst{1}, M. Kadler \inst{1} \and
       J.A. Zensus \inst{1} }

%   \offprints{T.G. Arshakian}
%   \mail{tigar@mpifr-bonn.mpg.de}

   \institute{Max-Planck-Institut f\"ur Radioastronomie, Auf dem
     H\"ugel 69, 53121 Bonn, Germany \email {tigar@mpifr-bonn.mpg.de}
     \and INAOE, AP 51 y 216, CP 72000, Puebla, Pue., Mexico
%\email {vahram@inaoep.mx}
     \and IA UNAM, AP 70-264, 04510 Mexico D.F., Mexico}

   \abstract{We combine Very-Long-Baseline Interferometry (VLBI) data
     for $\sim100$ active galactic nuclei (AGN) available from the
     Very Large Baseline Array (VLBA) 2\,cm imaging survey and optical
     spectroscopy to investigate the relationships in the
     emission-line region--central engine--radio jet system.
%relationship
%     between the parsec-scale radio jets, their central engines, and
%     the emission-line regions.
     Here, we present the diversity of spectral types among the
     brightest AGN in our sample. We also discuss correlations between
     the mass of the central engine and properties of the parsec-scale
     radio jet for 24 AGN selected by the presence of H$\beta$
     broad-emission lines in their spectra.

     \keywords{Galaxies: active -- black holes -- jets} } 
     \authorrunning{T.G. Arshakian et al.}
     \titlerunning{Radio-optical scrutiny of the central engine}
     \maketitle
%
%________________________________________________________________

\section{Introduction}
It is well known that supermassive central nuclei (or `black holes')
are responsible for activity in AGN producing the
bipolar relativistic jets of plasma material. We still don't know what
triggers the jet activity; the `magic' mechanism which transforms
the disk energy into the kinetic energy of the jet. To understand the
underlying physics it is crucial to study the black hole--jet
and black hole--disk couplings, looking for correlations between
physical and geometrical characteristics of the disk--black-hole--jet
system.
%OTHER WORKS ON THIS MATTER ...  IMPORTANCE OF PC-SCALE JETS ...

A wealth of information about the structure and kinematics of
parsec-scale jets is now available from the VLBA 2\,cm
survey\footnote{http://www.cv.nrao.edu/2cmsurvey} (Kellermann et
al. 1998, Zensus et al. 2002) and its statistically complete subsample
named the
%continuation known as the 
MOJAVE
\footnote{http://www.physics.purdue.edu/astro/MOJAVE}
survey. More than 200 core-dominated radio-loud AGN have been
monitored at 2\,cm (15\,GHz) wavelength since 1994 providing the
physical and detailed kinematical characteristics of compact core-jet
structures (Kellermann et al. 2004). Most of them have superluminal
features which implies that the jets are aligned at small angles to
the line of sight. To relate the properties of parsec-scale jets to
the properties of black holes and its optical environment, we combine
data available from the radio and optical domains (Arshakian et
al. 2004), i.e. from the VLBI monitoring and optical spectroscopy of
compact AGN which is capable of providing valuable information on the
properties of the central engine and the geometry and kinematics of
parsec-scale narrow-/broad-line regions (BLRs). Here, we present some
early results on inter-connections between the mass of black holes and
properties of parsec-scale jets in compact AGN.

%-------------------------------------------------------------
   \begin{figure}[h]
   \centering
   \includegraphics[width=0.5\textwidth]{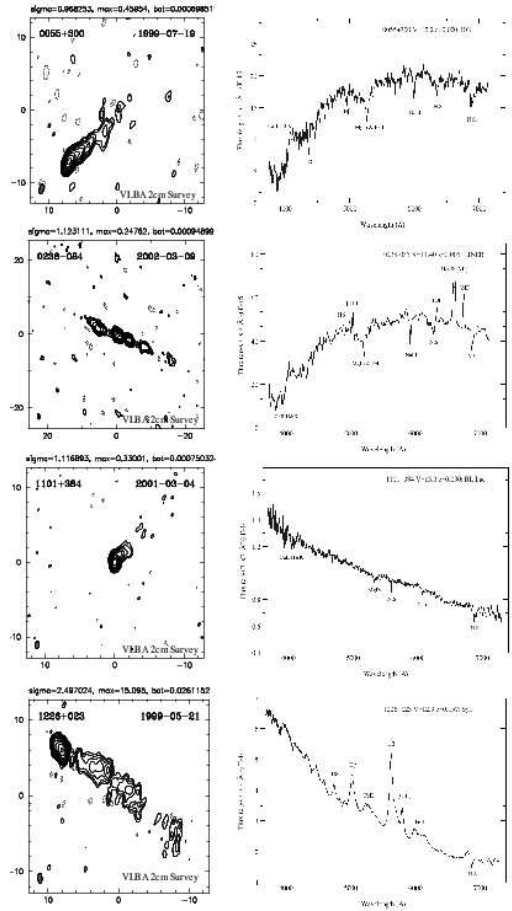}
      \caption{The VLBI images of selected AGN (left, axes are in
               milliarcseconds) combined with their optical
               spectra. Spectral classification shows diversity of AGN
               types: a radio galaxy (NGC\,315: 0055+300), LINER
               (NGC\,1052: 0238$-$085), BL Lac (Mrk\,421: 1101+384)
               and quasar (3C\,273: 1226+023).}
         \label{fig1}
   \end{figure}
%
%__________________

\section {The optical sample of compact AGN}
The 2\,m class optical telescopes GHAO (Cananea, Sonora, Mexico) and
OAN SPM (Baja California, Mexico) were used to carry out spectral
observations of optically bright AGN ($m<$ 18) from the VLBA 2\,cm
sample. So far, $\sim 70$ AGN have been observed using
intermediate-resolution spectroscopy (12\,\AA\, to 15\,\AA) with the
wavelength coverage $\sim$3800\,\AA\, to 8000\,\AA. High-resolution
spectra of $\sim$ 30 AGN were taken from the HST/SDSS archives and
Marziani et al. (2003). The spectral classification of $\sim 100$
radio sources showed a large diversity of AGN types: LINERs, Seyfert
galaxies, BL Lacs, quasars and radio galaxies (Fig.\ref{fig1}). The
complete, detailed list of all objects in the sample will be reported
in Arshakian et al. (in preparation). From this sample we selected 24
AGN having the H$\beta$ emission line in their spectra. 21 of the 24
AGN are type 1 (seven Seyfert 1 and 14 quasars), the remaining three
AGN are type 2 (two LINERS and one Seyfert 2). The 15 GHz radio
luminosity of 24 AGN varies over ($2\times10^{24}$ to
$2\times10^{28}$) W Hz$^{-1}$ over the redshift range 0 to 0.8.

We measured the width of the H$\beta$ broad emission lines and
luminosities of continua at 5100\,\AA\, and [O\,{\sc iii}] emission
lines. The broad H$\beta$ profile were prepared for measurements by
removing the narrow lines and Fe\,{\sc ii} blends. An empirical
Fe\,{\sc ii} template (V\'eron-Cetty et al. 2001) from I Zw 1 was
used. This template was broadened by convolution with Gaussian
profiles of constant velocity width and scaled to fit the broad
Fe\,{\sc ii} features at 4450\,\AA\, to 4700\,\AA\, and 5150\,\AA\, to
5350\,\AA. The subtraction of this template removes Fe\,{\sc ii} from
H$\beta$ and [O\,{\sc iii}] 5007 profiles. The [O\,{\sc iii}] $\lambda
5007$ profile was used as a template to remove narrow line component
of H$\beta$ from broad profile. The redshift of [O\,{\sc iii}]
$\lambda 5007$ was used to determine the wavelength of narrow
component of H$\beta$.

%                                                Two column figure
%----------------------------------------------------------- S_vib
   \begin{figure*}
   \centering
   \resizebox{\hsize}{!}{\includegraphics[clip=true]{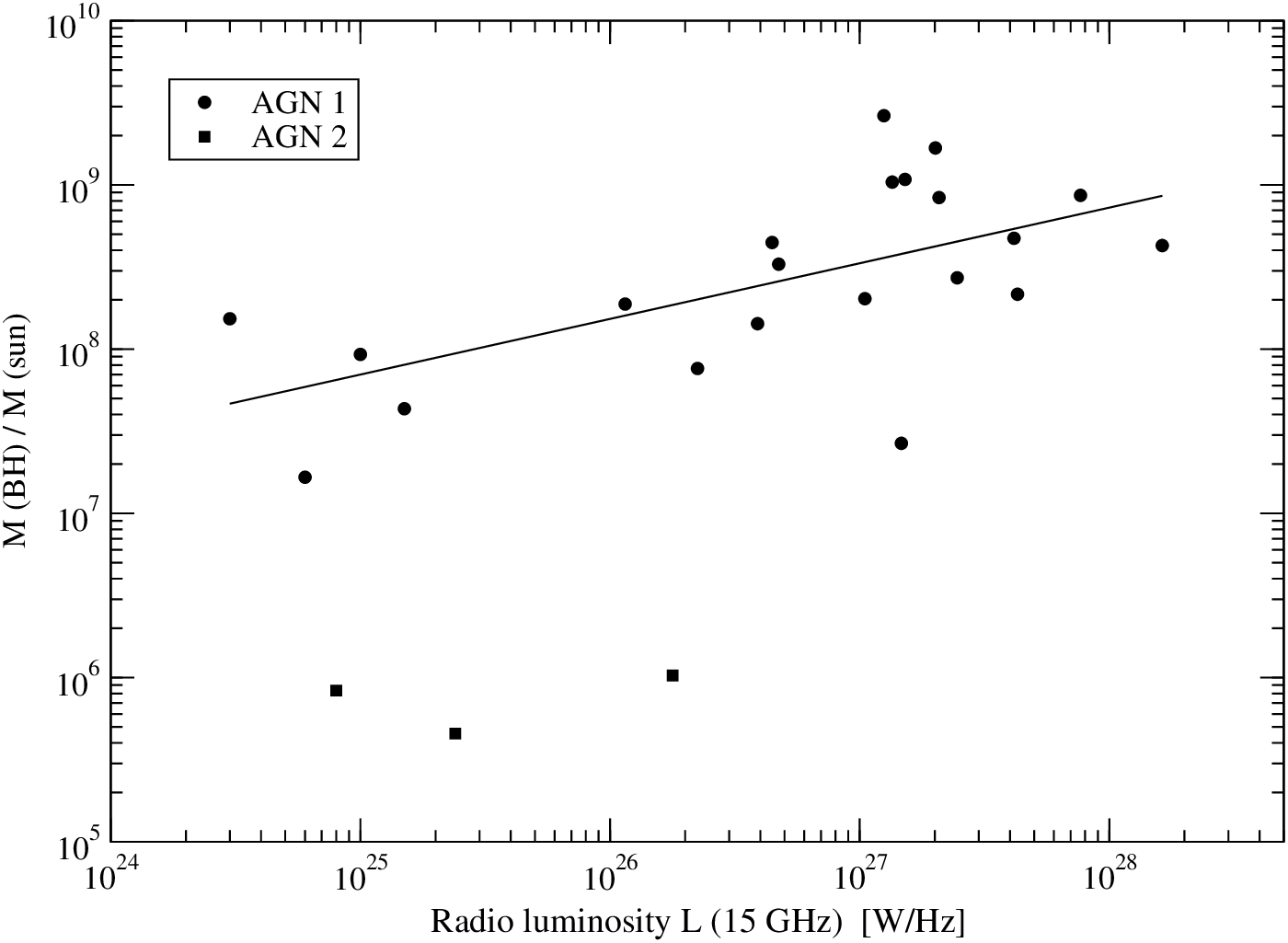}
   \includegraphics[clip=true]{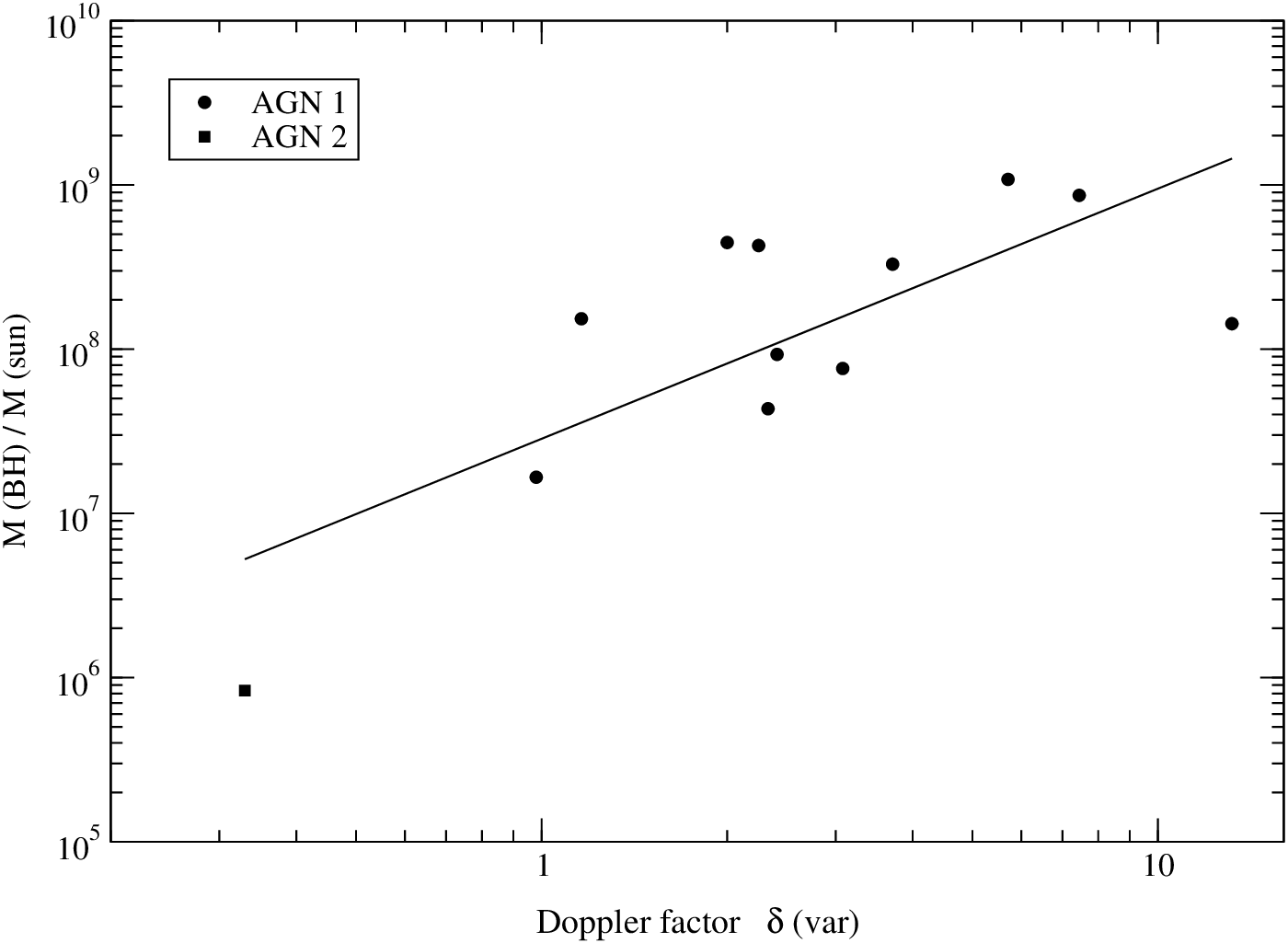}}
     \caption{The black hole mass versus radio luminosity at 15 GHz
     for 24 AGN (left panel), and the black hole mass versus Doppler
     factor (right panel) for 12 AGN with known estimates of a Doppler
     boosting determined from the radio flux density variability. Type 1
     AGN are marked by circles, and squares denote type 2 AGN.}
        \label{fig2}
    \end{figure*}
%
%______________________________________________________________
%
%                                                Two column figure
%----------------------------------------------------------- S_vib
%    \begin{figure*}
%    \centering
%    \resizebox{\hsize}{!}{\includegraphics[clip=true]{arshakianF3a.eps}
%    \includegraphics[clip=true]{arshakianF3b.eps}}
%      \caption{The black hole mass versus radio-loudness (left panel),
%      and apparent speed of the brightest component in the jet (right
%      panel). Designations are the same as in Fig.\ref{fig2}}
%         \label{fig3}
%     \end{figure*}
%______________________________________________________________

\section{Relations between $M_{\rm {BH}}$ and properties of the pc-scale radio jet}
The reverberation mapping technique (e.g. Kaspi et al. 2000) allows
the mass of black holes to be estimated from the width of H$\beta$
emission line and continuum luminosity at 5100\AA. The black hole mass
estimator works under the assumption that the motion of the broad
emission line region is virialised. We estimated the masses of
selected 21 radio-loud flat-spectrum type 1 AGN to be in the range
$M_{\rm BH}\sim (10^{7}$ to $3\times 10^{9})$ $M_{\odot}$ using the
reverberation mapping technique. This method underestimates the masses
of three type 2 AGN (Fig. 2, left panel) derived from narrow H$\beta$
emission lines.

The relation between $M_{\rm {BH}}$ and total radio luminosity at 15
GHz, $L_{\rm 15GHz}$, is shown in Fig. 2 (left panel). The positive
correlation for type 1 AGN is fitted by $L_{\rm 15GHz}\propto M_{\rm
BH}^{2.9\pm0.9}$ relation. Depending on the sample of flat-spectrum
quasars and the radio-loudness criterion used, the studies of the
$M_{\rm {BH}}-L_{\rm radio}$ relation for flat-spectrum quasars
produced contradictory results (Jarvis \& McLure 2002). The relation
$L_{\rm 5GHz}\propto M_{\rm BH}^{2.5}$ is suggested by Dunlop et
al. (2003) as an upper and lower envelope but separated by some five
orders of magnitude in radio power. Our result, the power-law index
$2.9\pm0.9$ is in agreement with 2.5 within $1\sigma$ error limit. The
apparent $L_{\rm 15GHz}$ of majority compact AGN are Doppler boosted,
$L_{\rm 15GHz}=\delta^{k}L_{\rm int}$, where $\delta$ is a Doppler
factor, $k$ is a constant and $L_{\rm 15GHz}$ is the intrinsic
luminosity. To understand the positive correlation in the $M_{\rm
BH}-L_{\rm 15GHz}$ plane, we test relations $M_{\rm BH}-L_{\rm int}$
and $M_{\rm BH}-\delta_{\rm var}$ for 12 AGN (Fig. 2, right panel)
with known Doppler factors $\delta_{\rm var}$ available from
L\"ahteenm\"aki \& Valtaoja (1999). No correlation is found with
intrinsic luminosity, while $M_{\rm BH}\propto\delta_{\rm
var}^{1.5\pm0.4}$. More massive central nuclei produce jets having a
high Doppler factor which is a function of the jet speed and jet
viewing angle. The positive $M_{\rm BH}-\delta_{\rm var}$ correlation
can be naturally explained if the mass of the black hole correlates
positively with the speed of the jet. More data are needed to confirm
this result.
%                                                Two column figure
%----------------------------------------------------------- S_vib
   \begin{figure}
   \centering
   \resizebox{6.5cm}{5cm}{\includegraphics[width=0.45\textwidth, bb
       =34 41 712 522, clip=true]{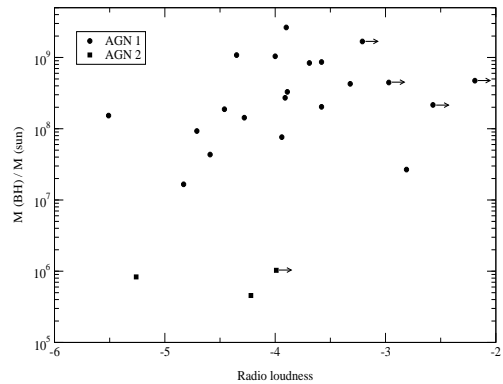}}
   \caption{The black hole mass versus radio-loudness. Designations
     are the same as in Fig.\ref{fig2}}
     \label{fig4}
   \end{figure}
%-------------------------------------------------------------
No clear dependence of the radio loudness (the ratio of the nuclear
luminosities at 5\,GHz and optical $B$ band) on $M_{\rm BH}$ is found
for type 1 AGN (e.g. Ho 2002). It is interesting to see how the black
hole mass correlates with radio-loudness of AGN in our sample. In
Fig. 3, the $M_{\rm BH}$ is plotted against
radio-loudness defined as the ratio of the radio flux density at 15
GHz to the X-ray flux between (2 and 10) keV (see Terashima \& Wilson
2003). Among radio-loud type 1 AGN, there is a positive trend of
increasing $M_{\rm BH}$ with radio-loudness, but a larger sample is
needed to confirm this correlation. No significant correlation is
found between $M_{\rm BH}$ and apparent speed of the brightest
component in the pc-scale jet.

Another positive correlation is found between $M_{\rm BH}$ and
[O\,{\sc iii}] emission-line luminosity (Fig.\ref{fig4}) in the form
$M_{\rm BH} \propto L_{\rm [O\,III]}^{0.76\pm0.14}$ ($\rho\sim
0.7$). The $ L_{\rm [O\,III]}$ is thought to be proportional to the
total photoionizing luminosity, $ L_{\rm tot}$, of the central engine
and/or shock waves from the jet. On the other hand the $ L_{\rm
[O\,III]}$ is a measure of the total kinetic power of the jet
(Rawlings \& Saunders 1991), $Q_{\rm jet} \propto L_{\rm tot} \propto
L_{\rm [O\,III]}$. It appears that the mass of the central engine in
compact AGN controls both the radiating power and the jet kinetic
power, $M_{\rm BH} \propto L_{\rm tot}^{0.76} \propto Q_{\rm
jet}^{0.76}$. If the $M_{\rm BH}\propto L_{\rm [O\,III]}^{0.76}$
correlation will stand for a larger sample of AGN then it can be used
for estimating the black hole masses directly from [O\,{\sc iii}]
emission-line luminosities.
%                                                Two column figure
%----------------------------------------------------------- S_vib
\begin{figure}
  \centering
  \resizebox{6.5cm}{5.1cm}{\includegraphics[width=0.45\textwidth, bb
      =-32 64 647 728, clip=true]{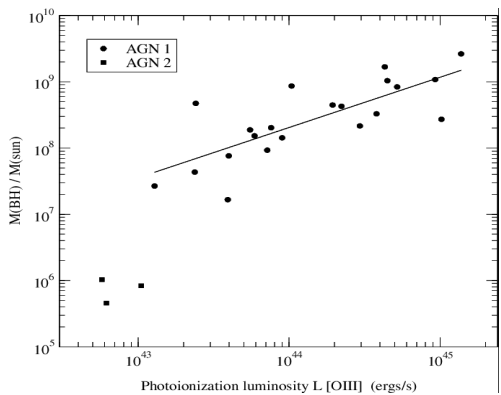}}
  \caption{The black hole mass against [O\,{\sc iii}]
    emission-line luminosity. The correlation is fitted with
    the $M_{\rm BH } \propto L_{\rm [O\,III]}^{0.76}$
    relation for 21 type 1 AGN.}
  \label{fig4}
\end{figure}
%-------------------------------------------------------------
%\section{Conclusions}

\begin{acknowledgements}
  We thank to M.-P. V\'eron-Cetty for kindly providing the Fe {\sc ii}
  template. Part of this work was done within the framework of the
  VLBA 2\,cm Survey collaboration. TGA and VHC acknowledge the CONACyT
  research grant 39560-F. M.\,K. was supported for this research
  through a stipend from the International Max Planck Research School
  (IMPRS) for Radio and Infrared Astronomy at the University of Bonn.
\end{acknowledgements}

\bibliographystyle{aa}

\end{document}